\documentclass[
  a4paper,
  twocolumn,
  longbibliography,
  aps,
  pra,
  superscriptaddress,
  amsfonts,amsmath,amssymb,
  floatfix
]{revtex4-2}
\usepackage[
  colorlinks,
  citecolor=blue,
  linkcolor=blue,
  urlcolor=blue
]{hyperref}
\usepackage{graphicx,epstopdf,bm}

\usepackage{braket}
\usepackage{blindtext}
\usepackage{enumitem}
\usepackage[capitalize]{cleveref}
\usepackage{xcolor}
\usepackage[all]{hypcap}
\usepackage{orcidlink}
\usepackage[separate-uncertainty = true]{siunitx}

\AddToHook{cmd/appendix/before}{\crefalias{section}{appendix}}
\crefname{section}{Sec.}{Secs.}
\Crefname{section}{Section}{Sections}
\crefname{appendix}{App.}{Appces.}
\Crefname{appendix}{Appendix}{Appendices}

\DeclareMathOperator{\Tr}{Tr}
\newcommand{\e}[1]{\mathrm{e}^{#1}}
\newcommand{\op}[1]{\hat{#1}}
\newcommand{\opdag}[1]{\hat{#1}^\dagger}

\newcommand{\trace}[2][]{\Tr_{#1}\left[#2\right]}

\newcommand{\deriv}[2][]{{\frac{\mathrm{d} #1}{\mathrm{d} #2}}}

\newcommand{\difd}{\mathrm{d}}
\newcommand{\hrho}[0]{\hat{\rho}}
\newcommand{\expval}[1]{{\langle#1\rangle}}

\newcommand{\cf}{cf.\ }
\newcommand{\eg}{e.g., }
\newcommand{\ie}{i.e., }

\newcommand{\figref}[2][]{Fig.~\hyperref[#2]{\ref*{#2}#1}}
\newcommand{\figureref}[2][]{Figure~\hyperref[#2]{\ref*{#2}#1}}

\newcommand{\opopenone}{\op{\openone}}

\newcommand{\changedBlock}[0]{\color{black}}
\newcommand{\changed}[1]{\textcolor{black}{#1}} %color for changes suggested by the referees

\begin{document}

\title{Obstacles to Continuous Quantum Error Correction via Parity Measurements}

%%% Authors %%%%%%%%%%%%%%%%%%%%%%%%%%%%%%%%%%%%%%%%%%%%%%%%%%%%%%%%%%%%%%%%%%%%

\author{Anton Halaski\,\orcidlink{0009-0001-9915-9470}}
\affiliation{Freie Universit\"{a}t Berlin, Fachbereich Physik and Dahlem Center for Complex Quantum Systems,
  Arnimallee 14, D-14195 Berlin, Germany
}

\author{Christiane P. Koch\,\orcidlink{0000-0001-6285-5766}}
\email{christiane.koch@fu-berlin.de}
\affiliation{Freie Universit\"{a}t Berlin, Fachbereich Physik and Dahlem Center for Complex Quantum Systems,
  Arnimallee 14, D-14195 Berlin, Germany
}

%%%%%%%%%%%%%%%%%%%%%%%%%%%%%%%%%%%%%%%%%%%%%%%%%%%%%%%%%%%%%%%%%%%%%%%%%%%%%%%%

\begin{abstract}
Time-continuous quantum error correction, necessary to protect quantum information under time-dependent Hamiltonians, relies on weak continuous syndrome measurements. Implementing these measurements requires a continuous coupling among at least two qubits and a meter, a demanding requirement. We show that, under continuous operation, common parity-measurement protocols in the circuit quantum electrodynamics platform corrupt the logical information. The failure arises from approximating the three-body interaction by a sum of two-body couplings to the meter, which prevents simultaneous suppression of measurement backaction on the logical and error subspaces. We argue that the same mechanism applies more generally beyond the circuit quantum electrodynamics setting. Taken together, our results impose a practical limitation on continuous stabilizer quantum error correction and point to the viable alternatives --- architectures that realize native three-body interactions, or erasure-based encodings in which the error subspace need not be protected.
\end{abstract}

%%%%%%%%%%%%%%%%%%%%%%%%%%%%%%%%%%%%%%%%%%%%%%%%%%%%%%%%%%%%%%%%%%%%%%%%%%%%%%%%

\maketitle

\section{Introduction}
  \label{sec:introduction}

Quantum error correction (QEC) is key to achieving fault tolerance in quantum computation~\cite{DevittRPP13,GirvinSciPostPhysLectNotes23}. In the circuit model, syndrome extraction is carried out with gate-based protocols~\cite{BluvsteinN24,GoogleQuantumAIandCollaboratorsN25,BrockN25}, 
assuming  that gates operate on timescales much shorter than the system's decoherence time. By contrast, computational approaches built around time-dependent Hamiltonians, such as adiabatic quantum computing~\cite{AlbashRMP18}, require a different approach to error correction. \changed{Since} time-dependent Hamiltonians generally do not commute with typical error operators~\cite{YoungPRX13,SarovarNJP13}, states that have suffered an error evolve differently from those that remain in the code space. \changed{This may} convert physical errors into logical faults. 

Weak continuous measurements \changed{may} mitigate this problem~\cite{Oreshkov_2013}: Their time-resolved records \changed{allow for estimating} when an error occurred \changed{such that the} unwanted evolution in the error subspace \changed{can be undone}. This is referred to as continuous QEC~\cite{Oreshkov_2013} \changed{and} requires, irrespective of the details of a given protocol, the ability to continuously measure the syndrome operators~\cite{AhnPRA02,MohseniniaQ20,AtalayaPRA21,BorahPRR22}. 
As a concrete \changed{and very simple} example, consider the three-qubit bit-flip code, where logical $\ket{0}$ and $\ket{1}$ are encoded as $\ket{000}$ and $\ket{111}$ and the syndromes are the pairwise parities $\op{Z}_1\op{Z}_2$ and $\op{Z}_2\op{Z}_3$. \changed{In} gate-based QEC, parity information \changed{is} transferred to a meter via subsequent two-body couplings. \changed{By contrast,} \textit{continuous} transfer of the parity information requires a three‑body coupling of the form $\op{Z}_i\op{Z}_{i+1}\op{M}$, where $\op{M}$ is some \changed{meter} operator. \changed{However,} hardware typically provides only two-body interactions, \changed{and} the three‑body term must be engineered. 

For transmon qubits, for example, a common realization \changed{couples} both qubits \changed{dispersively} to a readout resonator~\cite{LalumierePRA10,RisteN13}. \changed{This} yields an effective three‑body interaction \changed{but lifts the} degeneracy within one parity subspace. The resulting qubit–resonator entanglement, together with photon loss, produces effective dephasing~\cite{LalumierePRA10}. Proposed countermeasures include limiting the resonator photon number in the erroneous parity subspace~\cite{LalumierePRA10}, keeping track of the phase by accounting for all emitted photons~\cite{TornbergPRA10,FriskKockumPRA12}, or using a two‑photon drive with a nonlinear resonator to improve the signal‑to‑noise ratio~\cite{RoyerSA18}. All of these are very challenging to implement. Moreover, when the system switches from the parity subspace with large resonator occupation to the other subspace, a sudden change in photon number occurs which yields extra dephasing~\cite{LivingstonNC22}. 

Here we show that these types of measurement backaction are ubiquitous to measurement schemes based on two-body couplings. \changed{We find that backaction due to} non-equilibrium dynamics \changed{after a sudden change between} parity subspaces~\cite{LivingstonNC22} causes particularly large \changed{and unavoidable} errors. \changed{Moving beyond the example of the three-qubit bit-flip code and the platform of superconducting qubits, any} weak continuous syndrome measurement entails $N$-body interactions with $N>2$ that must be engineered \changed{from} native two‑body couplings. We show that for most choices of qubit encoding, continuous implementations that approximate these multi‑body couplings necessarily lift the degeneracy in at least one eigenspace of the syndrome operator. This introduces additional errors that seriously compromise the effectiveness of continuous QEC. We find that only erasure encodings~\cite{GrasslPRA97,KubicaPRX23} can be implemented using two-body couplings, since \changed{for this encoding} backaction in the error subspace \changed{does not affect} the logical information.

The paper is organized as follows. \Cref{sec:theoretical_framework} introduces the theoretical framework, starting with a brief review of digital and continuous QEC in \cref{subsec:intro_QEC}. Parity measurement protocols for superconducting qubits that allow for a continuous operation are summarized in \cref{subsec:review_parity_mnts}, together with the stochastic master equation to describe them. \Cref{sec:breakdown_protocols} illustrates  the problem of parity measurements in continuous QEC \changed{for the example of the three-qubit bit-flip code and} transmon qubits with parity readout via two transmission‑line resonators. We generalize these observations in \cref{sec:generalization}, showing  that the backaction is caused by engineering three‑body measurement terms from native two‑body couplings in \cref{subsec:2vs3body_general} and deriving conditions for faithful continuous parity measurements in \cref{subsec:requirements_3body}. Since these conditions imply significant hardware overhead, we suggest an alternative in \cref{sec:erasure_architecture}, that is, to implement continuous QEC with erasure qubits for which the measurements require only two‑body interactions. 
\Cref{sec:conclusion} summarizes and concludes.

%%%%%%%%%%%%%%%%%%%%%%%%%%%%%%%%%%%%%%%%%%%%%%%%%%%%%%%%%%%%%%%%%%%%%%%%%%%%%%%%
\section{Theoretical Framework}
  \label{sec:theoretical_framework}

\subsection{Syndrome measurements in QEC}
    \label{subsec:intro_QEC}

For the purpose of this work, it is sufficient to focus on one of the most simple QEC codes, the three-qubit repetition or bit-flip code~\cite{DevittRPP13}. The discussion proceeds analogously for phase flip codes and we allude to the implications for codes that correct against both bit flips and phase flips at the end of \cref{subsec:2vs3body_general}.
    
The three-qubit bit-flip code corrects for a single bit flip, i.e., an undesired Pauli-$X$ operation, by encoding a single logical qubit into three physical qubits. The two logical basis states are commonly chosen as 
\begin{equation}\label{eq:even-encoding}
       \ket{\overline{0}}=\ket{000}\,,\quad \ket{\overline{1}}=\ket{111}\,.    
\end{equation}
This is a stabilizer code~\cite{DevittRPP13} with two stabilizers, $\op{S}_1=\op{Z}_1\op{Z}_2$ and $\op{S}_2=\op{Z}_2\op{Z}_3$, where $\op{Z}_q$ is the Pauli-$Z$ operator acting on physical qubit $q$. Since $\op{S}_1$ and $\op{S}_2$ yield the parity of the two physical qubits they act on, they are also called parity operators. We refer to \cref{eq:even-encoding} as "even encoding" and to $\ket{\overline{0}}$ and $\ket{\overline{1}}$ as even-parity states.    
Another possibility is the "odd encoding",
\begin{equation}
        \label{eq:odd-encoding}
        \ket{\overline{0}}=\ket{010}\,,\quad\ket{\overline{1}}=\ket{101}\,,
\end{equation}
with corresponding stabilizers $\op{S}_{1}=-\op{Z}_1\op{Z}_2$ and $\op{S}_{2}=-\op{Z}_2\op{Z}_3$. We will see in \cref{sec:breakdown_protocols} that the choice of the logical subspace makes a difference for the implementation of the code. 

Conventional QEC based on the stabilizer formalism relies on repeatedly performing projective measurements of the stabilizers. The measurements should all yield $+1$ if the state is in the logical subspace. If, on one of the three physical qubits, a bit flip occurs, the state leaves the logical subspace, and measurement of at least one of the two stabilizers will yield a $-1$ eigenvalue. Since one can infer the qubit on which the bit flip has occurred from the combination  of outcomes, one can apply an $\op{X}_q$ correcting operation~\cite{DevittRPP13} to the faulty qubit.

In continuous QEC, projective measurements are replaced by continuous measurements of the stabilizers~\cite{AhnPRA02,AtalayaPRA21}. Instead of binary measurement outcomes $\op{S}_{1/2}^{(k)}\in\{+1,-1\}$ at discrete times $t_k$, time-continuous and real-valued measurement signals $\op{S}_{1/2}(t)\in\mathbb{R}$ are then collected~\cite{AtalayaPRA21}. While in an actual experiment, the signal would be averaged over some finite time interval $\Delta t$ and binned at discrete times $\op{S}_{1/2}(t_i)$~\cite{GuilminI23,Guilmin24}, the time discretization $\Delta t$ in the continuous measurement is still much finer than in the case of projective measurements. We assume the measurement signal to be time-continuous in the following and discuss the implications of finite time resolution in \cref{sec:breakdown_protocols}. 
In case of no error, the signals fluctuate around $+1$. Once an error occurs, one or both signals change to a mean value of $-1$, but this change takes time. Defining a syndrome uncertainty region $[\theta_l,\theta_u]$ with $-1<\theta_l<\theta_u<1$, only if both signals are outside of this region, a bit flip is detected with confidence and a correction operation applied~\cite{AtalayaPRA21}.
Assuming this operation to be instantaneous is justified as long as the time scale of the change of the Hamiltonian is much larger than the duration of the correcting gate~\cite{AtalayaPRA21}. In principle, the discrete feedback operation can also be replaced by a time-continuous one whose amplitude depends on the measurement signal~\cite{AhnPRA02} but implementing this in a practical protocol is even more challenging.

\subsection{Parity measurements in circuit QED}
\label{subsec:review_parity_mnts}

In order to measure their parity, two qubits need to interact with a meter system, changing the state of the meter, ideally in such a way that measuring the meter only yields information about the parity of the qubits. This is important because the meter introduces a backaction on the qubits, and any information about the population or relative phase within a parity subspace would result in dephasing of the logical state. 
In single-shot measurements (where the measurement signals are integrated over a short amount of time to obtain binary measurement outcomes), it is unlikely that errors occur during the measurement and the meter can be traced out, yielding an effective equation of motion for the qubits alone. By contrast, in continuous QEC, errors occur at random times leading to a potentially complex time evolution of the meter, and in turn a complex backaction on the qubits. It is thus necessary to treat the meter system explicitly.

Continuous measurements are commonly described using the framework of stochastic master equations~\cite{Wiseman10}. 
Taking the microwave resonator of circuit QED \changed{with annihilation operator $\op{a}$} as meter system and assuming homodyne detection \changed{of the $\op{\mathcal{Q}}=i(\opdag{a}-\op{a})$-quadrature~\cite{LalumierePRA10,TornbergPRA10,FriskKockumPRA12}, the} equation of motion reads ($\hbar\equiv1$)
\begin{align}
    \difd \hrho
    =& -i\left[\op{H}_m,\hrho\right]\difd t + \sum_{q=1}^{2} \gamma_q\mathcal{D}\bigl[\op{X}_q\bigr]\hrho\,\difd t\nonumber\\
    &+ \kappa\mathcal{D}[\op{a}]\hrho\,\difd t + \sqrt{\eta\kappa}\mathcal{H}\bigl[-i\op{a}\bigr]\hrho\,\difd W(t)\,,\label{eq:SME_parity}
\end{align}
where $\hrho(t)$ is the joint state of qubits and resonator (with time-dependence omitted for convenience in Eq.~\eqref{eq:SME_parity}), $\mathcal{D}[\op{c}]\hrho = \op{c}\hrho\opdag{c} - \frac{1}{2} \left\{\opdag{c}\op{c},\hrho\right\}$ and $\mathcal{H}[\op{c}]\hrho = \op{c}\hrho + \hrho\opdag{c} - \trace{\op{c}\hrho + \hrho\opdag{c}}\hrho$~\cite{Wiseman10}. The first term corresponds to the Hamiltonian evolution with the measurement Hamiltonian $\op{H}_m$, 
the second term describes bit-flip noise for each qubit at rate $\gamma_q$, and the third term corresponds to photon loss at a rate $\kappa$. The last term includes the stochastic Wiener increment $\difd W(t)$ and describes the information \changed{gained from the quadrature measurements with efficiency $\eta$}~\cite{Wiseman10}.  
The normalized homodyne current is given by~\cite{Wiseman10}
\begin{equation}
    I(t) = \changed{\expval{\op{\mathcal{Q}}}(t)} + \frac{1}{\sqrt{\kappa\eta}}\deriv[W(t)]{t}\,.\label{eq:homodyne_current}
\end{equation}
Finite time resolution in an experiment can be accounted for by a filter equation of the form~\cite{AtalayaPRA21}
\begin{equation}
    \dot{I}_\text{int}(t) = \frac{1}{\tau} [I(t) - I_\text{int}(t)]\,,
    \label{eq:filter}
\end{equation}
where $\tau$ is the averaging time and $I_\text{int}(t)$ the filtered signal. It is smoother than the direct signal $I(t)$ and thus easier to process. 

\begin{figure*}[tbp]
    \centering
    \includegraphics[width=\linewidth]{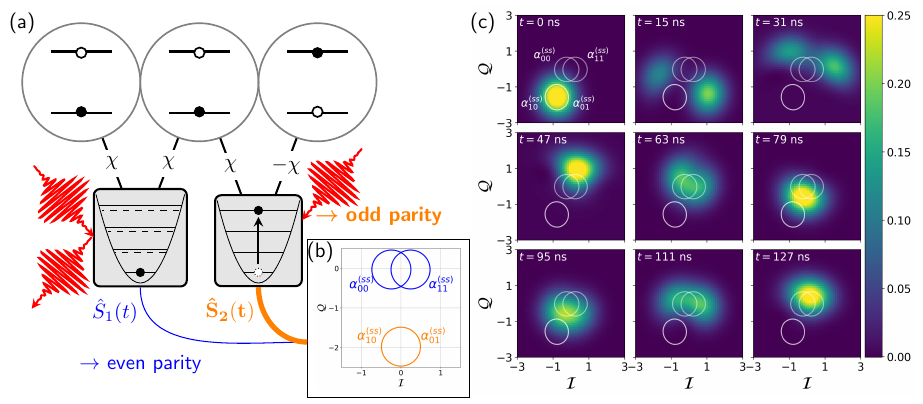}
    \caption{Continuous parity measurement scheme in the circuit QED architecture~\cite{LalumierePRA10}. (a) Sketch of three qubits (\eg transmons) (top) coupled to two readout resonators (middle) to measure the stabilizers of the three-qubit bit-flip code. The two left qubits are in an even-parity state which leads to a detuning of the resonator from the probe beam and a small homodyne signal. For the odd parity of the two right qubits, the dispersive shifts cancel and the drive is on resonance which leads to a large measurement signal. 
    (b) \changed{Contour lines corresponding to the resonator steady-state Wigner function taking the value of $0.25$, shown for the four computational qubit basis states.} 
    (c) Time evolution of the resonator Wigner function after an error occured: Starting from an equal superposition in the odd subspace, qubit 1 undergoes a bit flip at $t=0$, which initiates evolution towards the new (even) steady state. Due to higher-order terms in the Hamiltonian, the odd steady state is shifted to negative \changed{$\mathcal{I}$-values} compared to (b), see \cref{subsec:backaction_even_encoding} for details. %For better visualization, the bit-flip rate $\gamma$ is set to 0. 
    The parameter values for (b) and (c) are found in \cref{tab:parameters_figures_short}.}
    \label{fig:parity_mnt_summary}
\end{figure*}
For the purpose of our discussion, it is sufficient to focus on a single parity measurement.
% When measuring $\op{Z}_1\op{Z}_2$ for two superconducting qubits by coupling both of them dispersively to the same microwave resonator as in \cref{eq:H_disp_standard}, the system parameters are chosen in such a way that the effective couplings between the resonator and the two qubits are equal, $\chi_{1}=\chi_{2}=\chi$~\cite{LalumierePRA10}.
\changed{A possible way to measure $\op{Z}_1\op{Z}_2$ for two transmon qubits is to couple both of them dispersively to the same microwave resonator (see \cref{appendix:TIPT_results} for details). Furthermore, the system parameters must be chosen} in such a way that the effective coupling strengths between the resonator and the two qubits are equal, $\chi_{1}=\chi_{2}=\chi$~\cite{LalumierePRA10}. The Hamiltonian of the three-partite system in the frame rotating with the frequencies of qubits and resonator and after applying the rotating wave approximation \changed{can be approximated as}
\begin{equation}
    \op{H}_m = \mathcal{E}(\opdag{a} + \op{a}) + \chi(\op{Z}_1 + \op{Z}_2) \opdag{a}\op{a}\,.
    \label{eq:H_12r_measurement}
\end{equation}
In an experiment, the condition of equal couplings can be fulfilled by tuning the frequency of the qubits~\cite{LivingstonNC22}. In order to probe the resonator frequency shift induced by the qubits, a probe beam with amplitude $\mathcal{E}$ is applied at the dressed resonator frequency~\cite{LalumierePRA10,LivingstonNC22}, see also \figref[(a)]{fig:parity_mnt_summary}. 
If the two qubits are in the same state (even parity, left pair in \figref[(a)]{fig:parity_mnt_summary}), the resonator frequency is shifted by $\pm2\chi$, the drive is off-resonant and, as a consequence, the measurement signal small. By contrast, if the two qubits are in opposite states (odd parity, right pair in \figref[(a)]{fig:parity_mnt_summary}), the two frequency shifts cancel and the drive is on resonance, leading to a large measurement signal. It is thus possible to infer the parity from the measurement signal~\cite{LalumierePRA10}. This information can then be processed by a controller, which, in case of an error, applies a correction operation to the faulty qubit.
\begin{table}[tbp]
    \changedBlock
    \centering
    \caption{Parameters used in the simulations, % for the superconducting circuits. All parameters are 
    all given in units of $2\pi\times$\si{MHz}, of the same order of magnitude as in Ref.~\cite{LivingstonNC22}. A more detailed version including the device parameters is given in \cref{tab:parameters_figures_detailed} in \cref{appendix:TIPT_results}.}
    \begin{tabular}{c|r|r}
        & \figref[(c)]{fig:parity_mnt_summary} & Figs.~\hyperref[fig:parity_mnt_summary]{\ref*{fig:parity_mnt_summary}(b)}, \ref{fig:example_run_odd_and_even} \\\hline
        $\chi$      &  6.44 & 4.04 \\
        $\mathcal{E}$ & 2.10 & 1.40\\
        $\kappa$      & 3.00 & 2.00\\
        $\eta$        & 1.00 & 0.70\\
        $\gamma$      &    - & 0.02\\
        4th order terms? & yes & no
    \end{tabular}
    \label{tab:parameters_figures_short}
    \normalcolor
\end{table}
A first problem of this parity measurement scheme is considerable measurement backaction in the even subspace~\cite{LalumierePRA10,TornbergPRA10,FriskKockumPRA12}. The steady state of the driven resonator, when the qubits are in state $\ket{ij}$ (with $i,j\in\{0,1\}$), is well described by a coherent state $\ket{\alpha_{ij}^{(ss)}}$~\cite{LalumierePRA10} with the amplitude~\cite{GambettaPRA06}
\begin{equation}
    \alpha_{ij}^{(ss)} = \frac{-i\mathcal{E}}{\kappa/2 + i\chi_{ij}}\,,\label{eq:alpha_ij}
\end{equation}
where $\chi_{ij}=\bra{ij}\chi(\op{Z}_1 + \op{Z}_2)\ket{ij}$, \ie $2\chi,0,0,-2\chi$ for $\ket{00},\ket{01},\ket{10},\ket{11}$. The corresponding four steady states $\alpha_{ij}^{(ss)}$ are shown in \figref[(b)]{fig:parity_mnt_summary}. In the odd subspace, the measurement backaction is zero, provided the condition $\chi_{1}=\chi_{2}$ is fulfilled such that the two shifts cancel each other, and the resonator steady states are equal for $\ket{01}$ and $\ket{10}$. In the even subspace, however, the frequency shifts for $\ket{00}$ and $\ket{11}$ have opposite signs which leads to a finite separation of the two steady states.
As shown in \figref[(b)]{fig:parity_mnt_summary}, because the drive is detuned by the same absolute value, \changed{the average value $\expval{\op{\mathcal{Q}}}$ is equal, but $\expval{\op{\mathcal{I}}}=\expval{\opdag{a}+\op{a}}$ is different. Accordingly, a measurement of $\op{\mathcal{Q}}$} will yield information about the parity but none about the relative superposition of the states within a parity subspace. However, the different sign of the coupling terms leads to a different time evolution for the even-parity states. One can distinguish between two effects.

The first is an energy splitting between the two even-parity states, a direct consequence of the interaction term in the Hamiltonian, \cref{eq:H_12r_measurement}. The result is a relative phase shift between the two states, which we refer to in the following as "interaction-induced phase" $\overline{\phi}_e(t)$ \changed{(sometimes also referred to as "ac Stark shift"~\cite{LalumierePRA10})}. It grows as~\cite{TornbergPRA10}
\begin{align}
        \overline{\phi}_e(t) &= -4\chi\int_{t_0}^t \mathrm{Re}[\alpha_{00}(s)\alpha_{11}^*(s)] \difd s \nonumber\\
        &
        = 4\chi\int_{t_0}^t\mathrm{Re}[\alpha_{00}(s)^2] \difd s\,,\label{eq:phi_det}
\end{align}
where the identity $\alpha_{11}^*(t)=-\alpha_{00}(t)\ \forall t$ was used. Here, $t_0$ is some initial time, \eg the time that a bit flip occurred. In the steady state, $\alpha_{00}(t)$ is constant and this phase shift grows linearly in time, which makes it simple to correct. This becomes harder when considering non-equilibrium dynamics, as we show later.

The second effect is entanglement between resonator and qubits due to the time evolution of the resonator being conditional on the qubit state in the even subspace. The joint state then generally has the form $c_{00}\ket{00}\otimes\ket{\alpha_{00}} + c_{11}\ket{11}\otimes\ket{\alpha_{11}}$, which, after the loss of a photon, becomes proportional to $c_{00}\alpha_{00}\ket{00}\otimes\ket{\alpha_{00}} + c_{11}\alpha_{11}\ket{11}\otimes\ket{\alpha_{11}}$. When tracing out the meter, it becomes visible that the emission of light effectively changes the superposition of the even states. The result is a second kind of relative phase shift. It is the consequence of finite information about the even-parity states being stored in the meter, which is why we refer to it in the following as "information-induced phase". The phase grows with the distinguishability of the two resonator states, \ie with the difference between their real parts. The full equation reads~\cite{TornbergPRA10}
\begin{align}
    \tilde{\phi}_e(t)
    &= -\sqrt{\kappa \eta}\int_{t_0}^t \mathrm{Re}[\alpha_{00}(s) - \alpha_{11}(s)] \difd W(s)\nonumber\\
    &= -2\sqrt{\kappa \eta}\int_{t_0}^t \mathrm{Re}[\alpha_{00}(s)] \difd W(s)\,,
    \label{eq:phi_stoch}
\end{align}
where again $\mathrm{Re}[\alpha_{11}(t)]=-\mathrm{Re}[\alpha_{00}(t)]$ was used. The stochastic Wiener process $\difd W(t)$ appears because this phase shift depends on the photon statistics, \ie the measurement outcome. 

The \changed{phase shift} $\phi_e(t)=\overline{\phi}_e(t)+\tilde{\phi}_e(t)$ can in principle be estimated using \cref{eq:phi_det,eq:phi_stoch} and corrected~\cite{TornbergPRA10,FriskKockumPRA12}. This requires a perfect \changed{measurement chain} with $\eta=1$ to ensure knowledge of the full information-induced phase, cf.~\cref{eq:phi_stoch}, and, moreover, large bandwidth to detect all photons and reconstruct the full Wiener process $\difd W(t)$. If this is not possible, the phase information is lost over time and \changed{measurement-induced dephasing occurs~\footnote{Previous work~\cite{LalumierePRA10,LivingstonNC22} only considered the information-induced phase $\tilde{\phi}_e(t)$ as a cause of measurement-induced dephasing since the interaction-induced phase $\overline{\phi}_e(t)$ can be easily corrected when the system is in one of the steady states. In this work, we focus explicitly on the non-equilibrium dynamics and thus include both $\overline{\phi}_e(t)$ and $\tilde{\phi}_e(t)$.}.}
% Instead of recording the stochastic process, one can also reduce the steady state amplitude $\alpha_{00}^{(ss)}$ (resp.~$\alpha_{11}^{(ss)}$) to minimize the dephasing. Inserting \cref{eq:alpha_ij} into \cref{eq:phi_det,eq:phi_stoch}, the interaction-induced phase scales with $\mathcal{E}^2/\chi$ and the information-induced phase with $-\sqrt{\kappa\eta}\mathcal{E}/\chi$ in the limit $4\chi\gg\kappa$. This suggests to increase the effective interaction strength $\chi$ such that $\chi\gg\kappa,\mathcal{E}$ which corresponds to the drive being so off-resonant that the even steady state becomes effectively equal to the ground state of the resonator~\cite{LalumierePRA10,TornbergPRA10}. However, the Hamiltonian %\eqref{eq:H_disp_standard} 
% \eqref{eq:H_12r_measurement} is valid only in the limit $|g/\Delta|\ll1$, as explained in %\cref{subsec:circuit_QED}.
% \cref{subsec:review_parity_mnts}.
% Since $\chi$ scales with $g/\Delta$, there will be a limit on how large $\chi$ can be made. This will become even more important when using these parity measurement protocols in continuous QEC.

%%%%%%%%%%%%%%%%%%%%%%%%%%%%%%%%%%%%%%%%%%%%%%%%%%%%%%%%%%%%%%%%%%%%%%%%%%%%%%%%

\section{Measurement-induced errors in continuous QEC with superconducting qubits}
  \label{sec:breakdown_protocols}

The parity measurement protocols of Refs.~\cite{LalumierePRA10,TornbergPRA10,FriskKockumPRA12,RoyerSA18} were originally designed for single-shot measurements in digital quantum computing. For continuous operation, it is necessary to consider the resonator's non-equilibrium dynamics after a switch between subspaces and the possible impact of the measurement on the time evolution of the logical states (under some Hamiltonian $H(t)$ as in quantum simulation or adiabatic quantum computing). We show now that these factors cause the protocols to fail when used in continuous QEC. 
To this end, we shortly recall in \cref{subsec:odd-to-even_dephasing} the experimental findings of Ref.~\cite{LivingstonNC22} on strong dephasing for the odd encoding which is specific to continuous QEC. While it was conjectured in Ref.~\cite{LivingstonNC22} that the dephasing can be mitigated by using better hardware and more involved protocols, we show in \cref{subsec:dephasing_perfect_detector} that the problem persists for a perfect detector and, illustrating it with a simulation of the continuous QEC protocol, we attribute it to the unknown time at which the error occurs. In \cref{subsec:backaction_even_encoding}, we consider the measurement backaction in the case of even encoding, which has to the best of our knowledge, not yet been discussed. We conclude the section with a short review of alternative parity measurement protocols in \cref{subsec:beyond-disp}.

\changed{To support our arguments, we have solved \cref{eq:SME_parity} numerically for the measurement of a single parity, \ie a system of two transmon qubits coupled to one resonator.  We assume bit-flip noise on qubit $1$ with rate $\gamma_1=2\gamma$ (it is sufficient to model errors on only one qubit since the full three-qubit code identifies which qubit has flipped). %which is not possible with a single parity measurement. 
In the filtering equation \eqref{eq:filter}, we use $\tau=3/(\kappa\eta)$, which is approximately the optimal value determined in Ref.~\cite{AtalayaPRA21} for the parameter values we use. The simulations are performed using the Quantum Toolbox in Python (QuTiP)~\cite{Lambert24} with a time step of $\SI{0.4}{ns}/2\pi$ or smaller.}

\subsection{Odd-to-even dephasing}
  \label{subsec:odd-to-even_dephasing}

Encoding in the odd subspace, used in the continuous QEC experiment of Ref.~\cite{LivingstonNC22}, has the advantage of zero measurement backaction until an error induces a change to the even subspace where a phase shift is introduced. In addition to the dephasing discussed in \cref{subsec:review_parity_mnts}, a strong dephasing was observed~\cite{LivingstonNC22} when switching from the odd to the even subspace. This dephasing is due to the time-dependence of the resonator amplitude $\alpha_{00}(t)$ in \cref{eq:phi_det,eq:phi_stoch}. 

Since the resonator is initially in the odd steady state which has a large amplitude $\alpha_{01}^{(ss)}=\alpha_{10}^{(ss)}$, there is large entanglement and a large phase shift during the evolution from the odd to the even steady state. This is illustrated in \figref[(c)]{fig:parity_mnt_summary} showing the time evolution of the resonator Wigner function. Initially, the qubits are in an equal superposition of $\ket{10}$ and $\ket{01}$ in the odd subspace, and the corresponding two resonator states overlap. At $t=0$, a bit flip is induced and the qubits switch to the even subspace. Since $\mathrm{Re}[\alpha_{00}(t)^2]$ and $\mathrm{Re}[\alpha_{00}(t)]$ (which enter \cref{eq:phi_det,eq:phi_stoch}) remain large, a large phase shift occurs. In the simplest case when no effort is made to keep track of the phase during the measurement run, \ie when ignoring the information from the measurement outcome and setting $\eta=0$, then the coherence is reduced for a single deterministic bit flip by approximately a factor of $\exp\{-|\alpha_{01}^{(ss)}|^2\}$~\cite{LivingstonNC22} until the even steady state is reached. However, this description does not capture the core problem of odd encoding: The time at which the bit flip has occurred is unknown. \changed{This is discussed in Ref.~\cite{LivingstonNC22} in the context of static $ZZ$-interactions. In the following, we show that the unknown bit-flip time leads to strong dephasing even in case of a perfect detector and in absence of $ZZ$-coupling.}

\subsection{Odd-to-even dephasing despite a perfect detector}
  \label{subsec:dephasing_perfect_detector}

When collecting the measurement signal, it is in principle possible to estimate the \changed{phase shift $\phi_e(t)$} using \cref{eq:phi_det,eq:phi_stoch} and reduce the dephasing. 
Assuming a \changed{measurement chain} with efficiency $\eta$, the coherence is reduced only by approximately a factor $\exp\{-(1-\eta)|\alpha_{01}^{(ss)}|^2\}$, see \cref{appendix:coherence_odd_to_even}. This implies that, with a perfect detector, it is possible to fully estimate the phase of a single trajectory, avoiding the loss of logical information. 
In order to correct for the phase accumulated in the even subspace, the starting time $t_0$ in  both \cref{eq:phi_det,eq:phi_stoch} is needed. However, $t_0$ can only be roughly estimated from the measurement signal and the exact value is unknown. Further, the estimate of $t_0$ will be worse the smaller the measurement efficiency $\eta$. As a consequence, it is necessary to consider a range of times at which the bit flip could have occurred, \ie an ensemble of trajectories. Since the amplitude of the resonator state and correspondingly the derivative of the phase is initially large, the difference in phase between the different trajectories will also be large. In other words, even with a perfect detector with infinite bandwidth and $\eta=1$, there is uncorrectable dephasing.

\begin{figure}[tbp]
    \centering
    \includegraphics[width=\linewidth]{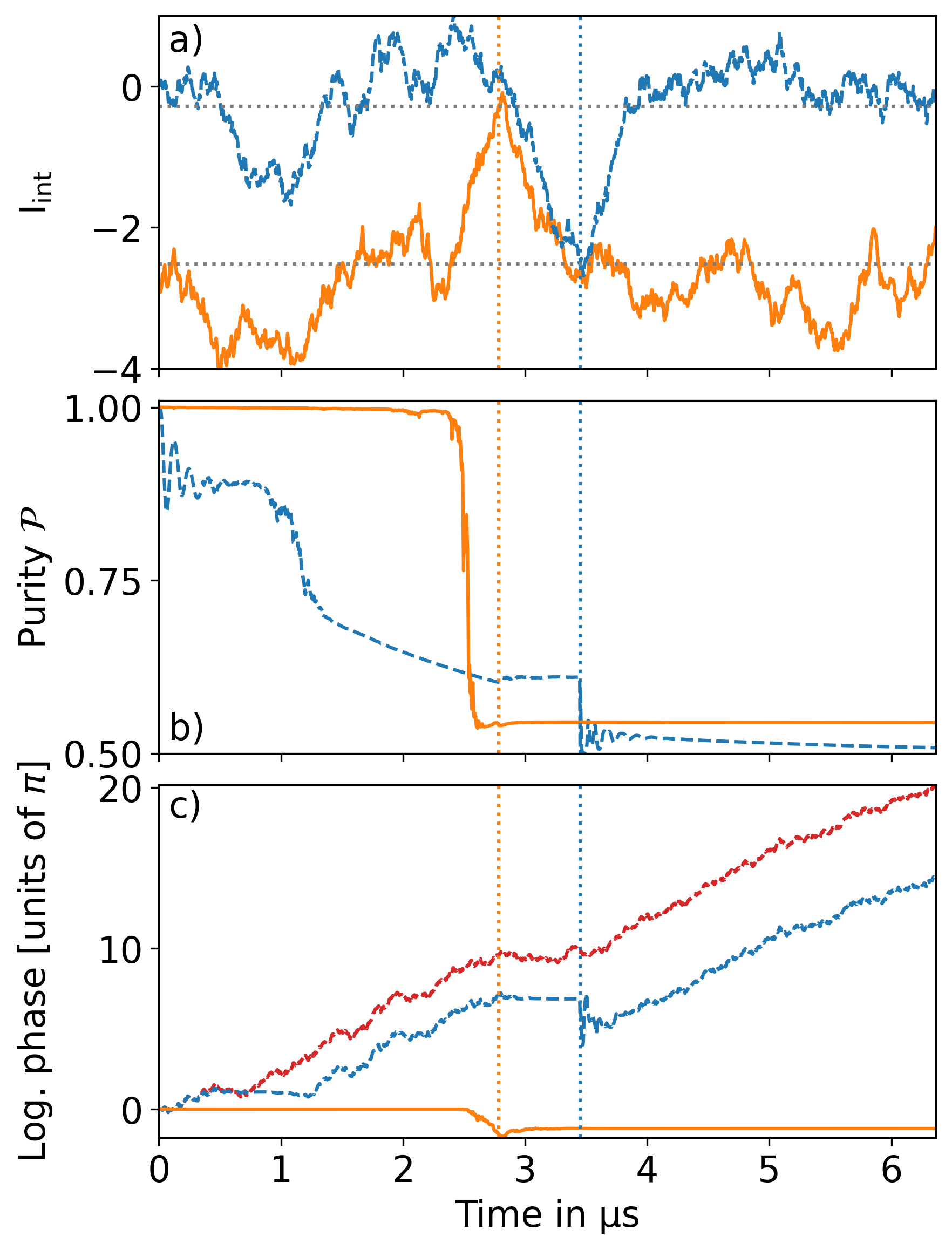}
    \caption{Example runs of the continuous QEC protocol where a bit-flip error was recorded for odd (solid, orange) and even encoding (dashed, blue). The initial state is $\ket{\overline{+}}=(\ket{\overline{0}} + \ket{\overline{1}})/\sqrt{2}$ where $\ket{\overline{0}/\overline{1}}$ are the the respective logical states for the given encoding. (a) Integrated photocurrent $I_\text{int}(t)$. Indicated are also the threshold values (dotted, gray) $\theta=(0.1 \overline{I}_o,0.9\overline{I}_o)$ with respect to the steady state value $\overline{I}_o=-2.8$ in the odd subspace. At the time the signal crosses the upper (lower) threshold for odd (even) encoding, here indicated by a vertical dotted orange (blue) line, an error is detected and a bit flip is applied to the faulty qubit by the controller. (b) Purity $\mathcal{P}$ (\cf \cref{eq:def_purity}) of the reduced state of the qubits in the respective subspace which contains the logical information, \ie the code space in case of no error or the respective error subspace in case of an error. (c) Logical phase (\cf \cref{eq:def_log_phase}) of the encoded logical information induced by the parity measurement. For even encoding, also the estimated phase (dashed, red), calculated using \cref{eq:phi_det,eq:phi_stoch}, is shown. The parameters are found in \cref{tab:parameters_figures_short}.}
    \label{fig:example_run_odd_and_even}
\end{figure}
The orange lines in \cref{fig:example_run_odd_and_even} display an example run of the continuous QEC protocol of Ref.~\cite{AtalayaPRA21} (with the thresholds shown as horizontal lines) using the parity measurement protocol of Ref.~\cite{LalumierePRA10}. The initial state of the qubits is taken to be $\ket{\overline{+}}_\text{odd}=(\ket{01} + \ket{10})/\sqrt{2}$ (which is maximally sensitive to phase shifts), and the resonator is initially in the ground state (but the initial transition from the ground state to the steady state is not shown for odd encoding). 
In case of no error, the filtered homodyne signal fluctuates around its mean value of $-2.8$. After a bit flip occurs, around \changed{$t=\SI{2.5}{\micro\s}$}, the signal evolves towards the mean value for the even steady state which is approximately zero. As soon as the signal hits the upper threshold (vertical orange line) around \changed{$t=\SI{2.8}{\micro\s}$}, the controller implements a bit flip on the faulty qubit, the qubits are back in the odd subspace and the signal evolves towards the steady state value of the odd subspace. Strong dephasing can be observed in \figref[(b)]{fig:example_run_odd_and_even} \changed{(orange solid line) where the purity of the logical state is shown. It is defined as
\begin{equation}
    \mathcal{P}^{(e/o)}(t) = \trace{\left(\frac{\op{P}_{e/o}\hrho_\text{L}(t)\op{P}_{e/o}}{\trace{\op{P}_{e/o}\hrho_\text{L}(t)}}\right)^2}\,,\label{eq:def_purity}
\end{equation}
where
\begin{equation}
    \hrho_\text{L}(t) = \trace[\text{res}]{\hrho(t)}\label{eq:def_rho_L}
\end{equation}
is the reduced density matrix obtained by tracing out the resonator, and $\op{P}_{e/o}$ the projector on the even/odd qubit subspace.} The purity drops around the time of the bit-flip error to about 0.55, which is close to the maximally mixed state for the logical qubit. While the graph shows results for a \changed{measurement chain} efficiency of $\eta=0.7$, this effect would also occur for $\eta=1$ with similar strength, with the purity dropping to about 0.63 for $\eta=1$. Crucially, increasing $\chi$ does not help since the derivative of the phase immediately after the bit-flip error scales with the steady-state amplitude in the odd subspace $\alpha_{01}^{(ss)}$ which only depends on $\kappa$ and $\mathcal{E}$, see \cref{eq:alpha_ij}. 
To summarize: In line with Ref.~\cite{LivingstonNC22}, we observe that large dephasing is introduced when switching from the odd to the even steady state. This dephasing is uncorrectable when using odd encoding due to the uncertainty when the bit flip occurs. We conclude that continuous QEC is fundamentally impossible when using this measurement scheme in combination with odd encoding, since a large part of the coherence of the logical state is lost after only one bit flip.

\subsection{Measurement backaction for even encoding}
  \label{subsec:backaction_even_encoding}

Given that there is a random phase shift for even encoding that needs correction or mitigation even when no error occurs, one may ask if this encoding is useful at all. 
\changed{Even encoding is not subject to the strong dephasing caused by non-equilibrium dynamics, which presents a fundamental issue for odd encoding.} We therefore quantify the resources required for full phase correction below and examine the consequences of measuring during a quantum computation. Further, we discuss now also the event of two subsequent bit flips which leads to an uncorrectable error, and explain why increasing the coupling strength between the meter and the qubits does not help to decrease the measurement backaction in the steady state.

For illustration purposes, we use again an example run of the continuous QEC protocol with the same parameters as in the previous section, but now for even encoding. This is shown in \cref{fig:example_run_odd_and_even} with the dashed blue lines. Analogously to the discussion for odd encoding above, \figref[(a)]{fig:example_run_odd_and_even} shows the integrated signal, with the steady state value now being (approximately) zero. 
\changed{Compared to the trajectory for odd encoding (orange solid lines), the dynamics is more complex: 
From $t=0$ to $t=\SI{0.5}{\micro\s}$, the resonator evolves from the ground to the even steady state. Then two bit flips, near $t=\SI{0.5}{\micro\s}$ and $t=\SI{1.2}{\micro\s}$,  occur in such rapid succession that they are not detected. Afterwards, the system undergoes equilibrium dynamics until $t=\SI{2.8}{\micro\s}$, where another bit flip occurs. This time, the error is detected once the signal hits the lower threshold value, around $t=\SI{3.4}{\micro\s}$ (vertical dotted blue line). An instantaneous correction operation is applied, followed by  equilibrium dynamics until the end of the simulation.}

\changed{We first estimate the resources for a real-time correction of the \changed{phase shift $\phi_e(t)$} under equilibrium dynamics, \ie without errors. The corresponding measurement backaction is seen} in \figref[(b,c)]{fig:example_run_odd_and_even} from \changed{$t=\SI{1.2}{\micro\s}$} to \changed{$t=\SI{2.8}{\micro\s}$} (where an error occurs). \changed{The logical phase in \figref[(c)]{fig:example_run_odd_and_even} is given by 
\begin{equation}
\begin{split}
    \phi_e(t) = \arg(\bra{00}\rho_\text{L}(t)\ket{11}),\\
    \phi_o(t) = \arg(\bra{01}\rho_\text{L}(t)\ket{10}),
\end{split}\label{eq:def_log_phase}
\end{equation}
for the even and odd subspace, respectively, with $\hrho_\text{L}$ defined in \cref{eq:def_rho_L}.
The phase shift} without any errors is evident in \figref[(c)]{fig:example_run_odd_and_even} in both the general upwards trend reflecting the interaction-induced phase $\overline{\phi}_e$ and a fluctuation corresponding to the information-induced phase $\tilde{\phi}_e$. 
The interaction-induced phase $\overline{\phi}_e$ in the steady state can in principle be estimated beforehand by performing measurements of the effective coupling strength $\chi$ and the steady-state amplitude $\alpha_{00}^{(ss)}$. The information-induced phase $\tilde{\phi}_e$, however, depends on the noise process and is random. It thus needs to be tracked during the whole measurement process, which leads to dephasing for an imperfect \changed{measurement chain}. This is seen in the continuous purity decrease in \figref[(b)]{fig:example_run_odd_and_even} from \changed{$t=\SI{1.2}{\micro\s}$} to \changed{$t=\SI{2.8}{\micro\s}$}. With realistic hardware, the measurement leads to slow dephasing of the logical state even without any errors.

Further errors arise from the non-equilibrium dynamics required to reach the respective steady state. It is triggered by the occurrence or correction of an error, and for even encoding also when the measurement protocol is started. On the one hand, similar to odd encoding, it is necessary to consider a mixture of possible trajectories when a bit flip is induced by the environment, where each trajectory now accumulates a different phase in the even subspace until the occurrence of the error. However, since the phase accumulation in the even steady state is small, the corresponding dephasing is negligible as seen by the small purity change at \changed{$t=\SI{2.8}{\micro\s}$} when the error occurs in \figref[(b)]{fig:example_run_odd_and_even}. On the other hand, although the time of the correction operation (\changed{$t=\SI{3.4}{\micro\s}$} in \figref{fig:example_run_odd_and_even}) is known, the rapid change of phase after the correction operation makes an accurate estimation from the measurement signal difficult, even with a near-perfect \changed{measurement chain}. This is visible in \figref[(b)]{fig:example_run_odd_and_even} by the purity dropping to almost  0.5 due to the finite detector efficiency ($\eta=0.7$ in the simulations). One could mitigate this dephasing by switching off the measurement drive for a short time or by inserting a displacement field of amplitude $-\alpha_{01}^{(ss)}$ into the resonator to bring the resonator to its ground state~\cite{HuembeliPRA17} before applying the correction to the faulty qubit. There will, however, always be finite non-equilibrium dynamics to approach the respective steady state, and this leads to dephasing. It can be seen, for example, in the purity drop from $t=0$ to \changed{$t=\SI{0.5}{\micro\s}$} in \figref[(b)]{fig:example_run_odd_and_even}, the time in which the resonator evolves from the ground state to the even steady state.

An uncorrectable logical error is introduced by two subsequent bit-flip errors, occurring near \changed{$t=\SI{0.5}{\micro\s}$} and \changed{$t=\SI{1.2}{\micro\s}$} in \cref{fig:example_run_odd_and_even}. These bit flips are not detected because the second one occurs before the filtered readout signal, changed by the first one, hits the detection threshold (lower horizontal line in \figref[(a)]{fig:example_run_odd_and_even}). Since the timing of the second bit flip (around \changed{$t=\SI{1.2}{\micro\s}$}) is unknown to the experimentalist, it leads to dephasing, as seen in the purity drop by about $0.2$. 
Moreover, the time spent in the odd subspace (in which the phase shift is zero) is not accounted for when estimating the phase shift in the even steady state. To illustrate this, \figref[(c)]{fig:example_run_odd_and_even} shows the estimated logical phase (dashed, red) under the assumption that no bit flips occur. It starts to deviate from the actual phase (blue) due to the time spent in the odd subspace from \changed{$t=\SI{0.5}{\micro\s}$} to \changed{$t=\SI{1.2}{\micro\s}$}. This additional dephasing could be reduced by a more elaborate estimation method~\cite{TornbergPRA10,ChenPRR20}, but these methods come with a large additional overhead.

Even if the \changed{phase shift $\phi_e(t)$} can be perfectly estimated, it may interfere with the operations implemented on the logical qubit and may lead to uncorrectable logical errors. This implies the necessity to correct the phase shift in real time. Further, the qubits are constantly entangled with the resonator in the even subspace. Since the implemented time evolution on the qubits does not affect the entanglement with the resonator, this leads to constant errors in the time evolution of the logical state.

\changed{Finally, measurement backaction could also be reduced by decreasing the steady state amplitude $\alpha_{00}^{(ss)}$ (resp.~$\alpha_{11}^{(ss)}$). This corresponds to the limit $\chi\gg\kappa,\mathcal{E}$ at which measurement backaction becomes zero, cf. \cref{eq:alpha_ij,eq:phi_det,eq:phi_stoch}, with the ratio $\chi/\kappa$ determining by how many linewidths the resonance frequency of the resonator shifts when changing between the even and odd subspace. While it is possible to obtain larger ratios by decreasing $\kappa$, this is limited since $\kappa$ also determines the decay time from one parity subspace to another and the measurement rate, \ie how fast bit flips are detected~\cite{LalumierePRA10,TornbergPRA10,LivingstonNC22}. 
Alternatively, the effective coupling strength could be increased. However, we have identified several additional problems with this approach:}
Increasing $\chi$ to a point at which the dephasing and phase shifts become negligible implies the quasi-dispersive~\cite{SchusterN07,GoerznQI17} or even strong-coupling regime~\cite{BlaisRMP21} for the meter-qubit coupling. Then both a qubit-state-dependent anharmonicity in the resonator and a three-body interaction of the form $\op{Z}_1\op{Z}_2\opdag{a}\op{a}$ appear, as discussed in detail in \cref{appendix:TIPT_results}. The anharmonicity squeezes the steady state, while the three-body interaction shifts it in \changed{$\mathcal{I}$}, as illustrated in \figref[(c)]{fig:parity_mnt_summary}. As a result, the resonator decay becomes asymmetric and the two states become distinguishable due to their different \changed{$\expval{\op{\mathcal{Q}}}$-values}, causing stochastic shifts of population. This is also visible in \figref[(c)]{fig:parity_mnt_summary} where the amplitude of one of the two states decreases, cf. the panel for \changed{$t=\SI{15}{ns}$}. Such problems occur also when adding a Kerr nonlinearity to the resonator and replacing the linear drive by a parametric two-photon drive~\cite{RoyerSA18}. Similar to the constraints on $\chi$, limitations also exist for the drive strength $\mathcal{E}$, respectively the resonator occupation, which governs the qubit relaxation rates and transition frequencies~\cite{PetrescuPRB20}.

\subsection{Beyond dispersive readout with a single resonator}
\label{subsec:beyond-disp}

Before generalizing our considerations to hardware beyond superconducting qubits, we show that the measurement backaction that is at the core of all the problems identified above persists when using more than one resonator for the parity readout. Coupling the qubits jointly to two microwave resonators and detecting the sum of their output fields allows for removing the distinguishability of the two even states during the equilibrium dynamics~\cite{DiVincenzoNJP13}. However, this protocol still suffers from measurement-induced errors during the non-equilibrium dynamics when relaxing to the even steady state~\cite{TornbergPRA14}. \changed{Alternatively, each qubit may be coupled to an individual, isolated cavity, with a probe beam subsequently scattered from these cavities, \eg via beam splitters, gaining a phase shift of approximately $0$ or $\pi$ depending on the qubit parity~\cite{KerckhoffPRA09}. However, this protocol suffers from the same problem as the one discussed in \cref{subsec:review_parity_mnts} -- finite distinguishability of the even states and resulting dephasing, with the exact same scaling in $\chi/\kappa$.}
%However, this protocol also suffers from the problem of finite distinguishability of the even states and resulting dephasing, which has the exact same scaling in $\chi/\kappa$ as the protocol discussed in \cref{subsec:review_parity_mnts}.}

%%%%%%%%%%%%%%%%%%%%%%%%%%%%%%%%%%%%%%%%%%%%%%%%%%%%%%%%%%%%%%%%%%%%%%%%%%%%%%%%

\section{General conditions for backaction-free continuous parity measurements}
  \label{sec:generalization}

In the previous section, we have shown that existing parity measurement protocols for the circuit QED architecture fail when used in continuous QEC. In Sec.~\ref{subsec:2vs3body_general}, we argue that this failure is not limited to the circuit QED architecture and identify the engineering of an effective three-body interaction as the origin of the problem. This insight points to possible remedies, which we discuss in Sec.~\ref{subsec:requirements_3body}.

\subsection{Failure of continuous parity measurements based on two-body interactions}
  \label{subsec:2vs3body_general}

The building blocks of circuit QED, few-level systems and harmonic oscillators, are ubiquitous in quantum computing. The few-level systems could also correspond to atoms, ions, or molecules, and the harmonic oscillators to quantized optical instead of microwave fields. This suggests that the problem of measurement backaction discussed in \cref{sec:breakdown_protocols} affects other, if not all quantum computing architectures. 
We use once more the three-qubit bit-flip code as paradigm for our discussion and then generalize to arbitrary QEC codes.

\begin{figure}[tbp]
    \centering
    \includegraphics[width=\linewidth]{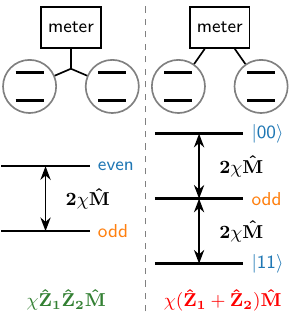}
    \caption{Parity measurements require three-body interactions. Left: For native three-body interactions, the states within each parity subspace are degenerate. Right: Typically native three-body interactions are not available and must be emulated by two two-body interactions. In this case, only the odd states are degenerate, whereas there is an energy splitting between the even states. This may lead to undesired entanglement with the meter as well as relative phase shifts or population differences.}
    \label{fig:level_scheme_2vs3body}
\end{figure}
As before, we focus on the continuous parity measurement of two qubits subject to bit-flip noise, for which the undesired measurement backaction arises from the difference in meter interaction for the two even-parity states $\ket{00}$, $\ket{11}$. This difference originates from the two-body qubit-meter interactions, as illustrated in \cref{fig:level_scheme_2vs3body}, where $\op{M}$ is an operator acting on the meter Hilbert space, and $\chi$ is the effective coupling strength between qubits and meter: For two-body interactions (right part of \cref{fig:level_scheme_2vs3body}), only the states in the odd subspace are degenerate, whereas an energy splitting of $4\chi\op{M}$ appears between the states in the even subspace~\changed{\footnote{The two qubits could have different coupling strengths $\chi_1$ and $\chi_2$, but this would make the situation only worse since then all four states have a different energy and are distinguishable. Changing the sign of one of the two couplings would swap the roles of even and odd subspace but not resolve the problem of non-degenerate states in one parity subspace.}.} 
Consequently, the even-parity states inevitably become distinguishable. By contrast, for a true three-body interaction of the form $\chi\op{Z}_1\op{Z}_2\op{M}$ (left part of \cref{fig:level_scheme_2vs3body}), all states within a given parity subspace are degenerate and thus indistinguishable.  

Continuous QEC proceeds by initializing the qubits in a logical basis state ("encoding"), followed by time evolution of the joint qubits-meter system, where the meter evolves differently depending on the qubit parity. Importantly, in continuous measurements, the readout rate is smaller than or at most of the same order as the rate of change of the meter state. As a consequence, while a change of qubit parity can happen quasi-instantaneously, the meter takes a finite amount of time to respond to this change, \changed{and non-equilibrium dynamics is guaranteed to occur. In the following, we identify this as the core issue of continuous parity measurements based on two-body interactions~\footnote{Our reasoning does not apply to autonomous QEC protocols where errors are corrected via engineered dissipation: With sufficiently large dissipation, the logical information always stays inside the same subspace such that there is no backaction due to non-equilibrium dynamics and thus no fundamental limit to time-continuous \textit{autonomous} QEC with the three-qubit bit-flip code~\cite{CohenPRA14}.}.} %maybe also \cite{KerckhoffPRL10}?

In order to show that measurements based on two-body interactions of the form $\chi(\op{Z}_1 + \op{Z}_2)\op{M}$ inevitably lead to backaction and dephasing of the logical states, let us consider possible meter realizations in general~\footnote{One may wonder whether the problem of two-body interactions can also be avoided when using an interaction of the form $\chi(\op{Z}_1 + \op{Z}_2)^2\op{M}$. However, this interaction contains the term $2\chi\op{Z}_1\op{Z}_2\op{M}$ and is thus equivalent to the assumption of a three-body interaction.}. 
In Sec.~\ref{sec:breakdown_protocols}, the meter is a resonator, i.e., a harmonic oscillator; alternatively, the meter could be an anharmonic $N$-level system. In this case, typically two levels are tuned to be resonant with the qubits, such that the multi-level nature of the meter is not crucial and to understand the basic principle of a measurement protocol it is sufficient to consider only two levels. 
In \cref{appendix:details_mnt_with_qubit}, we discuss the different measurement setups that are possible for such a "meter qubit". 
\changed{In general, the full Hamiltonian (neglecting time evolution on the qubits) in a frame rotating with the qubit frequencies reads
\begin{equation}
    \op{H}_m = \op{H}_0 + \chi(\op{Z}_1 + \op{Z}_2)\op{M}\,.
\end{equation}
Denoting the "free" meter Hamiltonian by $\op{H}_0$, at first glance, the two cases of  $[\op{M},\op{H}_0]=0$ and $[\op{M},\op{H}_0]\neq0$} (inducing an energy shift, respectively dynamics, on the meter) may seem to require separate treatments. Completing the description by accounting also for the meter readout provides, however, a unified perspective: In order to probe an energy shift continuously in time, one commonly measures the response of the meter to a drive with a given frequency, as in \cref{sec:breakdown_protocols}, but including the drive \changed{in $\op{H}_0$ results in $[\op{M},\op{H}_0]\neq0$ in both situations, see \cref{appendix:details_mnt_with_qubit} for an example.}

In the protocols in Sec.~\ref{sec:breakdown_protocols}, the meter approaches steady states for both parity subspaces (which must be different to distinguish the two parities), but one can also design measurement protocols in which only one of the parity subspaces is associated with a meter steady state, cf.~\cref{appendix:details_mnt_with_qubit}. In what follows we focus on protocols where the meter relaxes to a steady state for even parity, since protocols lacking this property exhibit persistent non‑equilibrium dynamics which makes them incompatible with continuous QEC, cf.~\cref{appendix:details_mnt_with_qubit}. In its approach to the steady state, the meter necessarily takes different paths for the two even-parity qubit states, irrespective of the protocol details. \changed{During this time, qubits and meter are entangled and the two even-parity qubit states $\ket{00}$, $\ket{11}$ accumulate different phases. For odd encoding, this results in uncorrectable errors since the exact transition time to the even subspace and thus the accumulated phases are unknown. For even encoding, this problem can be avoided as discussed in \cref{subsec:backaction_even_encoding}. 
%This may seem counter-intuitive at first glance since backaction occurs in the logical subspace whereas the (odd-parity) error subspace is backaction-free. In particular, this backaction implies an additional error source, \ie  entanglement and thus dephasing in the even steady state(s) which occurs when the meter states for $\ket{00}$, $\ket{11}$ are not equal. For example, in the protocol discussed in \cref{sec:breakdown_protocols}, the effective coupling strength enters linearly in the meter equation of motion and thus also in the denominator of the even steady state, cf.~\cref{eq:H_12r_measurement,eq:alpha_ij}. As a consequence, the meter steady states for the two even-parity qubit states become equal only in the limit of infinite coupling, which is unphysical (and already large coupling may introduce other problems as discussed in \cref{subsec:backaction_even_encoding}). 
We therefore focus on even encodings in the following. In this case, additional backaction may occur when the meter steady state is not equal for both even-parity states, cf. \cref{subsec:backaction_even_encoding}, but this source of backaction can be avoided with a nonlinear dependence on the effective coupling. This comes with more hardware requirements but is not a fundamental problem, cf. \cref{subsec:beyond-disp}.}

What remains as the essence of the backaction problem is the distinguishability of the even-parity states during the non-equilibrium dynamics of the meter, before it reaches the even steady state.
This dynamics is triggered by a bit flip, either caused by an error or a correction operation. A first consequence of this dynamics is entanglement between qubits and meter, which makes it impossible to continue applying logical operations from the moment the correction operation is carried out until the steady state is reached. Such disruptions of the computation is what continuous QEC had been designed to avoid, thus defeating its original purpose. Further, whenever two bit-flip errors occur in rapid succession, this leads to large uncorrectable errors, since the second bit flip causes a transition to the even subspace whose timing is unknown, see \cref{subsec:backaction_even_encoding}. 
In addition, depending on the details of the individual measurement protocol, there are further errors such as dephasing~\cite{LivingstonNC22}, or state collapse due to the two states becoming distinguishable by the parity measurement~\cite{TornbergPRA14}. In principle, these backaction effects could be avoided with a perfect detector and a real-time feedback controller with zero latency, but such hardware assumptions seem unrealistic. 
As a result, even these mainly practical issues represent a significant obstacle to the implementation of continuous parity measurements. 

Before disregarding measurement protocols that engineer a three-body interaction by two two-body interactions for continuous QEC altogether, one may wonder whether dynamical decoupling or similar techniques can mitigate the measurement backaction~\footnote{
One may also wonder whether the measurement strength can be made so large that the occurrence of errors is fully suppressed, similar to a quantum Zeno effect. When using odd encoding, the measurement backaction would then be zero at all times since the qubits never leave the logical subspace. However, the Zeno effect relies on the build-up of coherences between system and environment and thus cannot suppress Markovian noise~\cite{OreshkovPRA07}.}.
For example, a phase that is introduced to the measured qubits could be mitigated by flipping both qubits in periodic intervals~\cite{HuembeliPRA17}. However, these techniques generally work best when the phase derivative is independent of time which is not the case for the non-equilibrium dynamics during which most backaction effects occur. One could increase the switching speed to make it much faster than the timescale on which the state of the meter changes, but if the switching is performed too fast, the interaction in the even subspace will average out to zero, and even and odd subspace become indistinguishable. Furthermore, fast flipping of the qubits is in general not compatible with the time evolution by the logical Hamiltonian in analog quantum computing.

Importantly, our considerations carry over to other QEC codes which correct against phase flips or both bit flips and phase flips. First of all, our discussion also applies to continuous measurements of $\op{X}_1\op{X}_2$ which are necessary for a phase-flip code. This can be seen immediately by exchanging the $\op{Z}_q$-operators by $\op{X}_q$: Instead of dephasing, an interaction based on two-body couplings then leads to measurement backaction in the form of bit-flip errors. % or entanglement with the meter. 
Moreover, for both bit-flip and phase-flip codes, backaction becomes worse when measuring $N$-body syndromes with $N>2$. For example, when measuring $\op{Z}_1\op{Z}_2\op{Z}_3\op{Z}_4$ using two-body couplings to a meter, $(\op{Z}_1+\op{Z}_2+\op{Z}_3+\op{Z}_4)\op{M}$, both even and odd subspace become non-degenerate, with the even subspace splitting into three and the odd subspace splitting into two subspaces with different interaction Hamiltonians. As a result, there is measurement backaction in both logical and error subspaces.
\changed{The dephasing that is caused by parity measurements realized with two-body interactions can also not be mitigated when concatenating the bit-flip code with a phase-flip code: The detection of a bit-flip error would likely trigger a phase-flip error, whose detection would likely trigger another bit-flip error, resulting in a potentially endless loop of errors, which just increases the overall noise in the system.}
%Consider a fully correcting QEC code where all syndrome measurements are performed continuously, but are implemented via two-body interactions. The measurements targeting bit-flip errors then lead to dephasing of the joint state of all qubits involved in the measurements, at least at intermediate times. Due to the measurements targeting phase errors, this dephasing is transformed to a logical phase error with high probability. Moreover, these measurements are also faulty and may lead to bit-flip errors. 
%In short, a concatenation of bit-flip and phase-flip codes does not help to overcome the measurement backaction problem due to two-body qubit-meter interactions.

To summarize, irrespective of the details of a measurement protocol, when used in continuous QEC, two-body interactions between meter and qubits result in additional errors that defeat the purpose of error correction. Instead, true many-body interactions between meter and qubits are required, and we discuss in the following to what extent pure three-body interactions can be realized with current hardware.

\subsection{Experimental requirements for three-body interactions}
  \label{subsec:requirements_3body}

\begin{figure}[tbp]
    \centering
    \includegraphics[width=\linewidth]{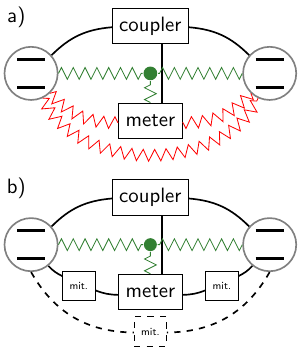}
    \caption{Is it possible to realize backaction-free continuous measurements when separating qubit-qubit coupling and qubit-meter coupling? (a) The necessary coupling between the qubits and the meter is mediated by a global coupler which, when integrated out, yields effective two-body (red) and three-body (green) interactions. (b) Additional local couplers can be used as mitigators (mit.) to address the two-body interactions which otherwise lead to undesired backaction. For the coupling scheme of Ref.~\cite{MenkePRL22}, the mitigator between the qubits would be optional because the two-body interaction is of $ZZ$-type, i.e, trivial for parity-encoded states.}
    \label{fig:general_coupling_scheme}
\end{figure}
Measuring a joint operator such as the parity of two qubits, $\op{Z}_1\op{Z}_2$, requires a coupling between the qubits themselves, in addition to their coupling to a meter. Instead of using a resonator as both meter and coupler as in the protocols discussed in \cref{sec:breakdown_protocols}, one can also separate these two roles. A possible configuration is shown in \figref[(a)]{fig:general_coupling_scheme}. An experimental implementation with superconducting qubits demonstrated the generation of an effective $ZZZ$-type three-body interaction, using magnetic fields to tune the coupling~\cite{MenkePRL22}. For the purpose of continuous QEC, it is, however,  not enough to realize a three-body interaction; one must also mitigate the detrimental effects due to the two-body terms that occur when switching between subspaces of different parity. But with physical two-body interactions, a three-body interaction can only emerge as a higher-order process of multiple two-body interactions such that the three-body terms are usually smaller than the two-body terms, (\cf\cref{appendix:TIPT_results}). In numbers, the three-body terms in Ref.~\cite{MenkePRL22} are of the order of a few MHz, compared to a few tens of MHz for the two-body terms, implying that the detrimental effects due to the two-body terms would dominate in continuous QEC. It will thus be necessary to extend the coupling scheme in \figref[(a)]{fig:general_coupling_scheme}, canceling at least the meter-qubit two-body terms. This was proposed in Ref.~\cite{MenkePRL22} and is shown in \figref[(b)]{fig:general_coupling_scheme}: By introducing additional couplers ("mitigators") between meter and qubits and tuning their parameters, it is possible to effectively remove the undesired two-body interactions. 
The resources to implement a pure three-body interaction without any two-body interactions between the meter and the qubits are thus a global coupler, which interacts, via two-body interaction, with the meter and both qubits, as well as one local coupler for each qubit, interacting with this qubit and the meter, to mitigate the two-body interactions. All of these couplers need to be frequency-tunable. The significant overhead of realizing a pure three-body interaction is likely the reason why, to the best of our knowledge, such measurement protocols have not been reported yet.

Once more, these considerations are not limited to superconducting qubits but can be generalized to other quantum hardware, due to lack of physical three-body interactions. For example, also in trapped ions, engineered three-body interactions are accompanied by two-body terms~\cite{BermudezPRA09,GambettaPRL20}. Moreover, these interactions as well as those between Rydberg atoms~\cite{FaoroNC15}, are typically limited to energy exchange processes which are not directly applicable to the non-demolition measurements necessary for continuous QEC. Therefore, whenever the fundamental physical interactions are local two-body terms, either between qubits, or between a qubit and a global degree of freedom, commonly a harmonic oscillator, backaction-free measurements for continuous QEC can only be realized with significant hardware overhead. Our reasoning may also be applied to bosonic codes where a continuous parity measurement requires high-weight terms in the interaction Hamiltonian, \ie large powers of the photon number operator~\cite{CohenPRL17}.

%%%%%%%%%%%%%%%%%%%%%%%%%%%%%%%%%%%%%%%%%%%%%%%%%%%%%%%%%%%%%%%%%%%%%%%%%%%%%%%%

\section{Continuous QEC with erasure detection}
  \label{sec:erasure_architecture}

\begin{figure}[tbp]
    \centering
    \includegraphics[width=\linewidth]{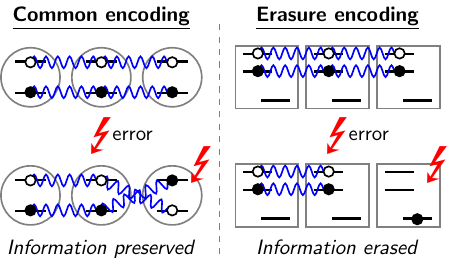}
    \caption{Basic idea of erasure encoding. For common encodings (left), the information or entanglement (here illustrated by wiggly lines) is preserved after an error, which makes it necessary to protect the code space and all error subspaces against measurement backaction. For erasure encoding (right), the information is erased by an error and only the code space must be protected against backaction.}
    \label{fig:erasure_encoding}
\end{figure}
We have shown so far that continuous QEC based on parity measurements with state-of-the-art hardware and standard encodings~\cite{DevittRPP13,GirvinSciPostPhysLectNotes23,Oreshkov_2013} is not feasible. This suggests to examine other encodings. A promising candidate is erasure detection~\cite{GrasslPRA97}. Erasure encoding has recently been realized in various platforms including superconducting qubits~\cite{KubicaPRX23,LevinePRX24,ChouNP24} as well as trapped atoms~\cite{WuNC22, SchollN23}, ions~\cite{KangPQ23}, and molecules~\cite{Holland24}. As schematically illustrated in \cref{fig:erasure_encoding}, erasure qubits are designed such that decay out of the code space (called erasure) is the dominant source of errors and all other errors can be neglected. 
Erasures work differently compared to common qubit encodings where two states of the physical qubit carriers make up both logical and error subspaces (left part of Fig.~\ref{fig:erasure_encoding}). In contrast, erasure qubits involve at least three states of the physical qubit carrier --- two states which are involved in forming the logical qubit and one that is populated only by the error.
An erasure, \ie a decay out of the code space destroys the entanglement between the faulty and the remaining qubits (right part of Fig.~\ref{fig:erasure_encoding}). This has two important consequences: On the one hand, the correction operation therefore needs to be entangling (just like the initial encoding operation), in contrast to standard encodings where it is a single-qubit gate. On the other hand, measurement of the erasure does not influence the information encoded in the remaining qubits. In other words, the measurement is inherently backaction-free. This is in contrast to Pauli errors such as bit or phase flips which preserve entanglement and for which a local measurement on the faulty qubit may introduce additional errors in the computational subspace, as discussed above. Moreover, detecting erasure errors, typically via population measurement in the erasure subspace~\cite{WuNC22}, only requires a two-body interaction between the faulty qubit and the meter. As a consequence, continuous erasure detection is readily achievable with state-of-the-art hardware. It does not incur an additional overhead, beyond encoding the erasure qubit itself. Erasure qubits are therefore the most promising and possibly the only realistic option for continuous QEC.

%%%%%%%%%%%%%%%%%%%%%%%%%%%%%%%%%%%%%%%%%%%%%%%%%%%%%%%%%%%%%%%%%%%%%%%%%%%%%%%%

\section{Conclusions and Outlook}
    \label{sec:conclusion}

We have examined protocols for continuous parity measurements in view of their usability in continuous QEC, identifying practical and fundamental problems for all protocols using standard encodings based on physical two-level system qubits. Taking the three-qubit bit-flip code as example and considering continuous parity measurements of superconducting qubits coupled to the same resonator, we have found large dephasing that arises from measuring the qubit parity via two two-body interactions between meter and qubits. 
While one can effectively engineer the three-body interaction required to measure a two-body operator such as parity with two-body interactions, this lifts the degeneracy within at least one parity subspace. 
Replacing the resonator by an arbitrary meter system, we have generalized our findings beyond the circuit QED architecture. Whenever the meter is used both to mediate the coupling between the qubits and to read out the state, we have found that the problem of inevitable dephasing of the logical states persists, irrespective of the choice of meter-qubit coupling and meter Hamiltonian. 
Our analysis is in line with the conjectured incompatibility of analog quantum computation and stabilizer QEC due to the necessity of many-body interactions~\cite{YoungPRX13,SarovarNJP13}.

A possible remedy is to use a tunable coupler in addition to the meter, to mediate the qubit-qubit coupling. This approach is not hampered by a fundamental obstacle, but incurs a significant hardware overhead since the desired three-body interaction is accompanied by undesired two-body interactions that need to be canceled by additional couplers ("mitigators"). A more practical alternative are erasure qubits which are inherently backaction-free and for which continuous syndrome extraction has already been demonstrated ~\cite{WuNC22}. While correction operations for erasures are entangling and thus more complex than in two-level-system-based encodings, the same operations are needed for the encoding step such that there is no overhead at the hardware level. Our findings thus suggest that the most practical way forward is to use erasure qubits not just in digital~\cite{BaranesPRX26} but also in continuous QEC~\changed{\cite{Elovenkova26}}.

%%%%%%%%%%%%%%%%%%%%%%%%%%%%%%%%%%%%%%%%%%%%%%%%%%%%%%%%%%%%%%%%%%%%%%%%%%%%%%%%

\section*{Acknowledgements}
We would like to thank Daniel Wennberg for help with the numerical solution of stochastic master equations at an early stage of this work. 
Funding from the Deutsche Forschungsgemeinschaft (DFG) – Project No. 277101999, Collaborative Research Centre (CRC) 183 (project C05)
is gratefully acknowledged.

\appendix
\section{Effective coupling Hamiltonian in the quasidispersive regime}\label{appendix:TIPT_results}

\changed{In the following, we derive the Hamiltonian of two transmons with base frequencies $\omega_q$ and anharmonicity $K_q$ ($q=1,2$), capacitively coupled to a single microwave resonator with frequency $\omega_r$ and coupling strength $g_q$ in the quasidispersive regime using fourth-order perturbation theory. The initial Hamiltonian reads (with $\hbar\equiv1$)~\cite{BlaisRMP21}}
%We start from \cref{eq:H_qr_base}, but extended to two qubits:
\begin{align}
    \op{H}_{qrq} = \omega_r \opdag{a}\op{a} + \sum_{q=1,2}\Bigl[\omega_q \opdag{b}_q\op{b}_q &- \frac{\changed{K_q}}{2} \opdag{b}_q\opdag{b}_q\op{b}_q\op{b}_q\nonumber\\
    &+ g_q(\opdag{a}\op{b}_q + \op{a}\opdag{b}_q)\Bigr].\label{eq:H_qrq_base}
\end{align}
\changed{where $\op{a}$ and $\op{b}_q$ are the annihilation operators of resonator and transmons, respectively, and the rotating wave approximation has been used.} 
To simplify notation, we define
\begin{align}
    \Delta_r &= \omega_r - \omega_1\,,\label{eq:Delta_r}\\
    \Delta_2 &= \omega_2 - \omega_1\,.\label{eq:Delta_2}
\end{align}
In the dispersive regime where $g_1/|\Delta_r|\ll1$ as well as $g_2/|\Delta_r-\Delta_2|=g_2/|\omega_r-\omega_2|\ll1$, one can treat the interaction term in the second line in \cref{eq:H_qrq_base} perturbatively and diagonalize the Hamiltonian. Since the interaction is off-diagonal with respect to the unperturbed Hamiltonian, all odd orders in $g_q$ vanish and the first non-vanishing order is the second. 
%When neglecting all higher orders, one arrives at \cref{eq:H_disp_standard} generalized to two transmon qubits. 
\changed{When neglecting all higher orders, one obtains the dispersive interaction in \cref{eq:H_12r_measurement}.}
There, processes involving both transmons play no role. 

To determine when the effect of higher-order terms become relevant, it is necessary to derive the next higher order, \ie the fourth order. We first sketch the derivation, which is based on the scheme in Ref.~\cite{ZhuPRB13}, and then discuss the implications of the different terms in the new Hamiltonian. We rewrite the perturbation term in the Hamiltonian,
\begin{alignat}{4}
    \op{V} 
    &\equiv& (\op{V}_1^+ &+ \op{V}_1^-)& &+& (\op{V}_2^+ &+ \op{V}_2^-)\nonumber\\
    &=&\, (g_1\op{b}_1\opdag{a} &+ g_1\opdag{b}_1\op{a})& &+& (g_2\op{b}_2\opdag{a} &+ g_2\opdag{b}_2\op{a}).\label{eq:def_V}
\end{alignat}
The index $q$ of $\op{V}_q^{\pm}$ indicates which transmon is involved in the exchange of excitations and $+ (-)$ indicates whether the resonator gains (loses) one excitation from (to) this transmon. It is crucial to treat the transmons including their higher levels (here simplified to anharmonic oscillators) because these are involved in the perturbation. Defining quantum numbers $(n,l_1,l_2)$ for the energy levels of the resonator and transmon 1 and 2, respectively, the fourth-order energy correction for the state $\ket{n, l_1, l_2}$ is given by
\begin{align}
    E_{N}^{(4)}
    = \sum_{M,P,Q\neq N} &\frac{V_{NM}V_{MP}V_{PQ}V_{QN}}{E_{NM}E_{NP}E_{NQ}}\nonumber\\
    &- \sum_{M\neq N} \frac{|V_{NM}|^2}{E_{NM}^2} \sum_{Q\neq N} \frac{|V_{NQ}|^2}{E_{NQ}},\label{eq:E_cor_4th-order}
\end{align}
where $N,M,P,Q$ are triple indices of the form $(n,l_1,l_2)$. $E_{NM} = E_N^{(0)} - E_M^{(0)}$ are the energy differences between the respective unperturbed eigenstates with eigenenergy $E_N^{(0)}$ for state $\ket{N}$, and $V_{NM}=\braket{N|V|M}$ are the corresponding matrix elements of the perturbation operator. This equation may seem intimidating, but we can break it down, treating the individual terms in \cref{eq:E_cor_4th-order} as paths in an energy diagram which involve four transfers of excitations. Using the fact that we only have to consider terms in which the initial and final state are the same, and making use of \cref{eq:def_V} reduces the sum to six combinations of perturbation operators. These are $\op{V}^+\op{V}^+\op{V}^-\op{V}^-$, $\op{V}^+\op{V}^-\op{V}^+\op{V}^-$, $\op{V}^+\op{V}^-\op{V}^-\op{V}^+$, and the same with $+$ and $-$ switched, where $\op{V}^\pm=\op{V}_1^\pm + \op{V}_2^\pm$. Reading the matrix elements from right to left, the first term describes for example the resonator first losing two excitations and then gaining them back. Imposing the constraint of equal final and initial state to the transmons means that only terms in which $\op{V}_q^+$ appears the same number of times as $\op{V}_q^-$ survive. This yields 6 remaining paths for each of the six combinations of $\op{V}^\pm$ resulting in a total of 36 possible paths. For the first combination, the paths are, \eg $\op{V}_1^+\op{V}_1^+\op{V}_1^-\op{V}_1^-$, $\op{V}_1^+\op{V}_2^+\op{V}_2^-\op{V}_1^-$, $\op{V}_1^+\op{V}_2^+\op{V}_1^-\op{V}_2^-$, and the same with 1 and 2 swapped. 

Next, we determine for each path the corresponding energy differences, and the matrix elements using \cref{eq:def_V}. Paths where the intermediate state after the second excitation transfer is the same as the initial state contribute to the second sum and paths where all three intermediate states are different contribute to the first sum, assuming that the uncoupled Hamiltonian is non-degenerate. 
Limiting the transmons to the first two levels and replacing $n \rightarrow\opdag{a}\op{a}$ and $l_q \rightarrow (\opopenone - \op{Z}_q)/2$ yields the full Hamiltonian in fourth-order perturbation theory,
\begin{align}
    \op{H}^{(4)}
    =&\; \tilde{\omega}_r\opdag{a}\op{a} + \changed{K_r}(\opdag{a}\op{a})^2 - \sum_{q=1,2}\frac{\tilde{\omega}_q}{2} \op{Z}_q\nonumber\\
    &+ \left[\chi_1 \op{Z}_1 + \chi_2\op{Z}_2\right]\opdag{a}\op{a} + \left[\nu_1\op{Z}_1 + \nu_2\op{Z}_2\right](\opdag{a}\op{a})^2\nonumber\\
    &+ \left[\zeta_{12} + \xi_{12}\opdag{a}\op{a}\right]\op{Z}_1\op{Z}_2.\label{eq:H_TIPT4_final}
\end{align}
The first line represents the transformed one-body terms, and the second and third line represent the coupling terms of transmons and resonator. As in second order, a correction to the dispersive shift appears, and $\chi_q=\chi_q^{(2)}+\chi_q^{(4)}$ is a combination of both second and fourth-order terms. In contrast to the second order, terms quadratic in the photon number appear, which lead to an effective anharmonicity of the resonator, where $K_r$ determines the qubit-independent and $\nu_q$ the qubit-dependent anharmonicity. Additionally, there is finite qubit-qubit coupling, so-called $ZZ$-interaction, determined by the parameter $\zeta_{12}$. Lastly, a three-body term with coupling strength $\xi_{12}$ appears. Note that a term of the form $(\opdag{a}\op{a})^2\op{Z}_1\op{Z}_2$ does not appear because the corresponding prefactor is exactly zero.
The parameters used in the different simulations are summarized in \cref{tab:parameters_figures_detailed}.
\begin{table}[tbp]
    \centering
    \caption{Parameters used in the simulations for the superconducting circuits. All parameters are given in units of \changed{$2\pi\times$\si{MHz}}. Above the double line are the physical parameters of the actual device \changed{defined in \cref{eq:H_qrq_base,eq:Delta_r,eq:Delta_2}}, and below are the parameters of Hamiltonian \eqref{eq:H_TIPT4_final} which are derived from these. The fourth-order corrections to the parameters for Figs.~\hyperref[fig:parity_mnt_summary]{\ref*{fig:parity_mnt_summary}(b)} and \ref{fig:example_run_odd_and_even} were artificially set to zero to show that the protocol even breaks down when neglecting higher-order effects. The parameters are of the same order of magnitude as the ones used in Ref.~\cite{LivingstonNC22}, \ie close to physical reality. Since only frequency differences influence the simulations, the exact values for the resonance frequencies are arbitrary and can be fixed by choosing for example \changed{$\omega_r/2\pi=\SI{6000}{MHz}$}, which is a typical value for the microwave fields used in superconducting architectures.}
    \begin{tabular}{c|rr}
         & \figref[(c)]{fig:parity_mnt_summary} & Figs.~\hyperref[fig:parity_mnt_summary]{\ref*{fig:parity_mnt_summary}(b)}, \ref{fig:example_run_odd_and_even} \\\hline
        $g_1$         & 94.84  &   89.00 \\
        $g_2$         & 72.46  &  108.77 \\
        \changed{$K_1$}    & 232.27 &  286.55 \\
        \changed{$K_2$}    & 383.80 &  304.79 \\
        $\Delta_r$    & 420.00 &  620.00 \\
        $\Delta_2$    &  62.66 & -184.78 \\\hline\hline
        $\chi_1$      &  6.44 & 4.04 \\
        $\chi_2$      &  6.44 & 4.04 \\
        $\nu_1$       & -0.20 & - \\
        $\nu_2$       & -0.20 & - \\
        \changed{$K_r$}    & -0.04 & - \\
        $\zeta_{12}$  &  0.23 & - \\
        $\xi_{12}$    & -0.47 & - \\
        $\eta$        & 1.00 & 0.70 \\
        $\mathcal{E}$ & 2.10 & 1.40\\
        $\kappa$      & 3.00 & 2.00\\
        $\gamma$      &    - & 0.02\\
    \end{tabular}
    \label{tab:parameters_figures_detailed}
\end{table}

We now discuss the implications of the higher-order terms on the parity measurement protocol in more detail, starting with the three-body term $\opdag{a}\op{a}\op{Z}_1\op{Z}_2$. Interestingly, this is actually the term that would be necessary for a perfect parity measurement, \ie without additional backaction on the qubits, as explained in \cref{subsec:2vs3body_general}. One problem is that the fourth-order effects are commonly much weaker than the second-order effects and have to be seen as a perturbation of the parity measurement protocol. For the parameters in the first column of \cref{tab:parameters_figures_detailed}, $\xi_{12}$ is more than one order of magnitude smaller than $\chi_{1/2}$, which means that this coupling is not helpful but actually harmful for the protocol of Ref.~\cite{LalumierePRA10}. The reason is the following. In this protocol, the frequency of the readout drive is chosen to be equal to the resonance frequency in the odd subspace where the resonator is undetuned such that the absolute value of the detuning is the same for both states in the even subspace. The three-body term now shifts the resonance frequency of the resonator in the odd (even) subspace by $-\xi_{12}$ ($+\xi_{12}$). Then the drive is not centered between the frequency of both states in the even subspace anymore but shifted by $\xi_{12}$ towards one of them. This makes the two states distinguishable which would not only lead to dephasing but to a full collapse of the initial qubit state into one of the two even states. To counteract this, the drive frequency must not be on resonance with $\tilde{\omega}_r$ but detuned by $\xi_{12}$. But this means that the drive is not fully on resonance in the odd subspace anymore but detuned by an amount of $2\xi_{12}$. This effectively leads to a lower signal-to-noise ratio for the same driving strength and to a shift of the odd steady state to finite values of the \changed{$\op{\mathcal{I}}$-quadrature} as depicted in \figref[(c)]{fig:parity_mnt_summary}. The implications of this are discussed in the main text. 

Second, anharmonicities are induced in the resonator, which leads to a shift and small squeezing of the odd steady state, visible in \figref[(c)]{fig:parity_mnt_summary}. The most noticeable effect is, however, that another necessary condition is introduced for indistinguishability of the states in the odd subspace, namely not only $\chi_1=\chi_2$, but also $\nu_1=\nu_2$. Otherwise, the two odd states would be distinguishable. This condition cannot be easily fulfilled and thus severely limits the available parameter space. Therefore, we have chosen a value of $\Delta_r$ and optimized the remaining five system parameters with the Nelder-Mead method until both conditions were met. (For the parameters of \cref{fig:example_run_odd_and_even}, we have only optimized for the condition $\chi_1=\chi_2$ because we have used the Hamiltonian in second-order perturbation theory.) 

Lastly, the pure ZZ-coupling given by $\zeta_{12}$ is not very relevant for the three-qubit bit-flip code because $\op{Z}_1\op{Z}_2$ is a stabilizer of the code and thus only introduces trivial operations. For other encodings, this may become relevant.

%%%%%%%%%%%%%%%%%%%%%%%%%%%%%%%%%%%%%%%%%%%%%%%%%%%%%%%%%%%%%%%%%%%%%%%%%%%%%%%%

\section{Loss of coherence for a single odd-to-even decay event}\label{appendix:coherence_odd_to_even}

When taking into account the information that is contained inside the measurement signal, the dephasing during the transition from odd to even steady state can be reduced. We follow the derivation in the supplementary material of Ref.~\cite{LivingstonNC22} but use the effective dephasing rate $\Gamma_\phi(t) = 4\chi\mathrm{Im}[\alpha_{00}(t)^2] - 2\kappa\eta\mathrm{Re}[\alpha_{00}(t)]^2$~\cite{TornbergPRA10}, which takes into account the information from the measurement signal. When assuming for simplicity that the effect of the drive is negligible in the even subspace due to the detuning, then the time-dependent coherent state amplitude is $\alpha_{00}(t)=\alpha_0\exp[-kt]$ with $k=\kappa/2 + 2i\chi$ and the initial amplitude $\alpha_0=\alpha_{01}^{(ss)} = -2i\mathcal{E}/\kappa$, the steady state amplitude in the odd subspace. The coherence is then reduced during the decay by a factor $\exp[-\zeta]$. Since $\alpha_0$ is fully imaginary, $\alpha_0^2 =-|\alpha_0|^2$ and $\mathrm{Re}[\alpha_0z] = |\alpha_0|\mathrm{Im}[z]$ for any complex number $z\in\mathbb{C}$. Then,
\begin{align*}
    \zeta
    &= \int_0^\infty \Gamma_\phi(t) \difd t\\
    &= |\alpha_0|^2\int_0^\infty \left\{-4\chi\mathrm{Im}[\e{-2kt}] - 2\kappa\eta\mathrm{Im}[\e{-kt}]^2\right\}\difd t\\
    &= |\alpha_0|^2\int_0^\infty \e{-\kappa t}\left\{-4\chi\mathrm{Im}[\e{-4i\chi t}] - 2\kappa\eta\mathrm{Im}[\e{-2i\chi t}]^2\right\}\difd t\\
    &= |\alpha_0|^2\int_0^\infty \e{-\kappa t}\left\{4\chi\sin(4\chi t) - 2\kappa\eta\sin^2(2\chi t)\right\}\difd t\,.
\end{align*}
The second term in the integral can be transformed using trigonometric identities and integration by parts as
\begin{align*}
    \eta&\int_0^\infty (-\kappa)\e{-\kappa t}2\sin^2(2\chi t)\difd t\\
    &= \eta\int_0^\infty (-\kappa)\e{-\kappa t}(1-\cos(4\chi t))\difd t\\
    &= \eta\e{-\kappa t}(1-\cos(4\chi t))\biggr|_0^\infty - 4\eta\chi\int_0^\infty\e{-\kappa t}\sin(4\chi t)\difd t \\
    &= - 4\eta\chi\int_0^\infty\e{-\kappa t}\sin(4\chi t)\difd t\,,\\
\end{align*}
which is exactly the negative of the first term in the integral, except for a factor of $\eta$. The result of the integral is%~\cite{LivingstonNC22}
\begin{align}
    \zeta &= 4\chi(1-\eta)|\alpha_0|^2 \int_0^\infty\e{-\kappa t}\sin(4\chi t)\difd t\nonumber\\
    &= -4\chi(1-\eta)|\alpha_0|^2 \mathrm{Im}\left[\int_0^\infty\e{-2k t}\difd t\right]\nonumber\\
    &= -4\chi(1-\eta)|\alpha_0|^2 \mathrm{Im}\left[\frac{1}{2k}\right]\nonumber\\
    &= (1-\eta)|\alpha_0|^2 \frac{16\chi^2}{\kappa^2 + 16\chi^2} \approx (1-\eta)|\alpha_0|^2\,,
\end{align}
under the assumption that $4\chi\gg\kappa$. This means that the coherence of the logical state is reduced by a factor of $\exp\{-(1-\eta)|\alpha_{01}^{(ss)}|^2\}$, which is the result in the main text.

%%%%%%%%%%%%%%%%%%%%%%%%%%%%%%%%%%%%%%%%%%%%%%%%%%%%%%%%%%%%%%%%%%%%%%%%%%%%%%%%

\section{Detailed discussion of parity measurements with a meter qubit}\label{appendix:details_mnt_with_qubit}

\changed{We discuss here the different possible measurement setups for a system consisting of two data qubits coupled to another qubit which comprises the meter. In a frame rotating with the data qubit frequencies, the Hamiltonian has the general form
\begin{equation}
    \op{H} 
    = -\frac{\omega_m}{2}\op{Z}_m + \chi(\op{Z}_1+\op{Z}_2)\op{M} + \op{H}_d(t)\label{eq:def_H_for_qubit_meter}
\end{equation}
with meter frequency $\omega_m$, effective coupling strength $\chi$, the coupling operator $\op{M}$ acting on the meter, and a possible measurement drive $\op{H}_d(t)$. The measurement observable is denoted by $\op{O}$ with corresponding measurement signal $I(t)$. For the sake of clarity, we treat the two cases $[\op{M},\op{Z}_m]=0$ and $[\op{M},\op{Z}_m]\neq0$ separately.}

\changed{We start with the case $[\op{M},\op{Z}_m]=0$, which corresponds to $\op{M}=\op{Z}_m$, or to an effective frequency shift on the meter. We assume that this shift is probed spectroscopically, \ie via measuring the response to a monochromatic drive with strength $\mathcal{E}$ and frequency $\omega_d$, 
\begin{equation}
    \op{H}_d(t)= \mathcal{E}\cos(\omega_dt)\op{X}_m\,.
\end{equation}}%
Since the measurement should not yield any information about the relative superposition in the even subspace, the response of the meter has to be the same for both even states. Due to the low complexity of the meter system, the only sensible choice for the drive frequency \changed{$\omega_d$} is to be equal to the meter frequency \changed{$\omega_m$} in the odd subspace (the line labeled "odd" in the right part of \cref{fig:level_scheme_2vs3body}) such that the frequencies for the even subspace lie symmetric around it. This means that the meter undergoes resonant Rabi oscillations in the odd subspace and off-resonant ones in the even subspace, and the difference can be measured to infer the parity, for example by detecting the light emitted from the meter via homodyne detection. This procedure is the protocol discussed in \cref{sec:breakdown_protocols}, adapted to a qubit meter where the drive in the frame rotating with the meter frequency is \changed{approximately $\op{H}_d=\mathcal{E}\op{X}_m/2$}. Unfortunately, this also means that similar problems occur. In particular, only in the limit of infinite coupling strength is the excited state population in the even steady state fully zero. For finite coupling, the meter has finite energy in the steady state and the two even states are different which leads to entanglement and dephasing. For non-equilibrium dynamics, the dephasing effect is even larger.

One may wonder whether dephasing can be avoided when measuring $\op{O}=\op{Z}_m$ or $\op{O}=\op{Y}_m$ via a coherent interaction with a second meter, \ie when the information is obtained in a different way. In the language of stochastic master equations, the measurement-induced action on the meter is described by the collapse operator. \Cref{eq:SME_parity} describes the direct detection of emitted light, corresponding to a collapse operator $-i\sqrt{\kappa}\op{a}$. For a qubit, the equivalent collapse operator would be $-i\sqrt{\kappa} \op{\sigma}^-$. Here instead we discuss $\sqrt{\kappa}\op{Y}_m$ (resp.~$\sqrt{\kappa}\op{Z}_m$) in the following. While the former always leads to a collapse to the ground state, the latter leads to a collapse into one of the $\op{Y}_m$-eigenstates (resp.~$Z_m$-eigenstates). Both even states are indistinguishable because they have the same $Y$-component respectively $Z$-component for all times. But there are other problems. For a $\op{Y}_m$-measurement, the meter undergoes rotations around an axis which lies in the $XZ$-plane for both even and odd subspace. As a consequence, the average $Y$-value is always zero and both parities become indistinguishable by the measurement. 
For a $\op{Z}_m$-measurement, the situation is almost equal to the one in \cref{sec:breakdown_protocols}, only that now both eigenstates of $\op{Z}_m$ are stabilized. Similarly, odd-to-even dephasing and finite steady-state dephasing in the even subspace occur. Additionally, finite fluctuations of the meter state can induce a flip to the other $\op{Z}_m$-eigenstate in the even subspace, even without a bit flip on the qubits. During these transitions, the variance in the $\expval{\op{Z}_m}$-value is large, leading again to strong dephasing for an imperfect detector. 
To summarize, an interaction based on an energy shift of the meter qubit always leads to finite entanglement and dephasing.

\changed{We now turn to the second case where $[\op{M},\op{Z}_m]\neq0$, \ie the coupling drives transitions in the meter. As a consequence, an additional measurement drive is not necessary, and we set $\op{H}_d(t)=0$. Typically, the energy splitting of the meter $\omega_m$ is much larger than the coupling strength $\chi$, implying that the effect of the qubit-meter coupling would be barely noticeable. A possible remedy is to modulate the coupling strength $\chi$ at a frequency $\omega_c$ which is equal or close to the transition frequency $\omega_m$ of the meter~\cite{DidierPRL15}. Then, one can go into a rotating frame in which the coupling becomes time-independent. For maximum distinguishability of even and odd parity, the detuning of meter and modulation frequency should be zero, $\omega_m=\omega_c$. When choosing (without loss of generality) $\op{M}=\op{X}_m$, the full Hamiltonian in the rotating wave approximation reads
\begin{equation}
    \op{H} = \chi(\op{Z}_1+\op{Z}_2)\op{X}_m\,.
\end{equation}}%
We first discuss a weak measurement of $\op{O}=\op{Z}_m$. In fact, a measurement protocol of this kind was proposed in Ref.~\cite{ChenPRR20}, albeit under the assumption that a three-body interaction can be realized.
At the beginning of the measurement, the meter is initialized in its ground state $\ket{0}_m$. In the odd subspace, the interaction is zero and the measurement current $I(t)$ fluctuates around $+1$, the ground state eigenvalue. The measurement stabilizes the system in $\ket{0}_m$ since it leads to a collapse towards the $\op{Z}_m$-eigenstates. Since the interaction between qubits and meter is zero, there is also no backaction in the odd subspace. 
In the even subspace, the qubits induce Rabi cycling in the meter qubit with a Hamiltonian of $\pm2\chi\op{X}_m$, with a different sign for $\ket{00}$ and $\ket{11}$ corresponding to a rotation of the meter anti-clockwise or clockwise around the $X$-axis. If the measurement rate is smaller than the frequency of the Rabi oscillation, the meter oscillates between ground and excited state which leads to a fluctuation of the measurement signal $I(t)$ around $0$ when averaging over a period. The two parity subspaces can thus be distinguished via the measurement signal. From this discussion, one can conclude that only $\op{O}=\op{Z}_m$ is a sensible choice since the two even states are distinguishable under a measurement of $\op{Y}_m$, and the $X$-component of the state is always zero, yielding no information about the parity in case of a measurement of $\op{X}_m$.

In contrast to the qubits coupling to the energy of the meter, there is no continuous phase shift in the even subspace because $\expval{\op{M}}(t)=\expval{\op{X}_m}(t)=0\ \forall t$ (the meter starts in a state with zero $X$-component and only rotates in the $ZY$-plane). 
Assuming an interaction of the form $\chi(\op{Z}_1 + \op{Z}_2)\op{X}_m$ and an initial state $\ket{\overline{+}}\otimes\ket{0}_m$ with $\ket{\overline{\pm}}=(\ket{00}\pm\ket{11})/\sqrt{2}$, the joint state at any time can be written as $\cos\phi\ket{\overline{+}}\otimes\ket{0}_m -i\sin\phi\ket{\overline{-}}\otimes\ket{1}_m$, $\phi\in[0,\pi)$. This illustrates the issue of this measurement protocol. The qubits become periodically entangled with the meter with the entanglement being the greatest when the meter state is on the equator, and the phase of the qubits is flipped when the meter is in state $\ket{1}_m$. Due to the entanglement, and since the measurement outcome on the meter is stochastic, the phase of the logical state in the even subspace during the measurement is effectively undetermined. This makes it impossible to apply this readout scheme when time evolution is implemented on the qubits. 
It is further not clear how the $\op{Z}_m$-measurement is performed. In general, to measure a certain observable continuously, this observable also needs to be part of the interaction with the readout system. For example, we have seen before that in order to perfectly measure $\op{Z}_1\op{Z}_2$, an interaction of the form $\op{Z}_1\op{Z}_2\op{M}$ is necessary. Similar to this case, in order to measure the $Z$-component of the meter, an interaction involving $\op{Z}_m$ would be necessary. In turn, this means that there will be possible backaction on the meter by the measurement. Because $\op{Z}_m$ is measured, the interaction can induce an effective $Z$-drift, which in turn would lead to $\expval{\op{M}}(t)=\expval{\op{X}_m}(t)\neq0$, \ie a phase shift on the qubits. A detuning between meter frequency and the modulation frequency of the interaction, \changed{\ie $\omega_m\neq\omega_c$}, has the same effect. 
To summarize, an interaction which drives transitions in the meter qubit leads periodically to strong entanglement between qubits and meter in the even subspace, such that no logical operations can be performed on the qubits. This justifies the assumption in the main text that the meter decays to a steady state in case of even parity. If the meter instead underwent coherent oscillations, there are two possibilities. Either the oscillations occur in a subspace of the meter which is unaffected by the interaction with the qubits and the oscillations are thus irrelevant for the discussion, or they occur in a subspace which is affected by the interaction, which means that the oscillations are different for the two even states. As explained above, this leads to the periodic buildup of entanglement between meter and qubits as long as the latter are in the even subspace.

%%%%%%%%%%%%%%%%%%%%%%%%%%%%%%%%%%%%%%%%%%%%%%%%%%%%%%%%%%%%%%%%%%%%%%%%%%%%%%%%

\nocite{CohenPRA14}
\nocite{OreshkovPRA07}
\bibliography{references}

\end{document}